# Second law versus variation principles

W.D. Bauer, email: W.D.BAUER@t-online.de

*Abstract:*

The field-dependent equilibrium thermodynamics is derived with two methods: either by using the potential formalism or by the statistical method. Therefore, Pontrjagin's extremum principle of control theory is applied to an extended ensemble average. This approach allows to derive the grand partition function of thermodynamics as a result of a control problem with the Hamilton energy. Furthermore, the maximum entropy principle follows and thereby the second law in a modified form. The derivation can predict second law violations if cycles with irreversibilities in varying potential fields are included into consideration. This conclusion is supported indirectly by experimental data from literature. As an example the upper maximum gain efficiency of a cycle using the polymer solution polystyrene in cyclohexane as dielectrics is estimated to less than 1 promille per cycle.

**Note added in proof 28$^{th}$ October 2003:** Comparing this preprint work with an analogous ferrofluidic system discrepancies are is found which show that the concrete model proposed here in section 4 is insufficient to settle the question. A way to solve the problem is proposed.

*1. Introduction*

The big success of the second law of thermodynamics relies on the fact that it predicts the direction of the known irreversible processes correctly. The inherent problem with it is that it is based on experience. Therefore, due to the axiomatic character of second law the question arises incidentally, whether the second law is an overgeneralisation.

On the other hand unconsciously and without any notice, other basic concepts are used sometimes in order to explain the direction of irreversibilities in thermodynamics. This can be the case if a chemist speaks about that his reaction is "driven by enthalpy".

Landau and Lifshitz [1] obtain the direction of electro-thermodynamic irreversibilities by the application of variational principles on potentials.

Because variational principles are included in the mathematics of a physical problem, the question arises, whether the second law as additional physical principle becomes obsolete if this purely mathematic aspect is included completely into consideration .

This article derives a field dependent equilibrium thermodynamic and checks the consistence and equivalence of the second law against a purely mathematical approach using the variational principles applied to potentials.

## 2. The derivation of the thermodynamic formalism including potential fields

The total Hamilton energy $H^*$ of a thermodynamic system including a potential field U is

$$\begin{aligned} H^* &= H(S(\vec{r}),\vec{r},n_i(\vec{r})) + U(S(\vec{r}),\vec{r},n_i(\vec{r})) \\ &= H(S(\vec{r}),\vec{r},n_i(\vec{r})) + \sum_i \int U_i(S(\vec{r}),\vec{r},n_i(\vec{r}))\, dn_i(\vec{r}) \end{aligned} \qquad (1)$$

where $H$ is the Hamilton energy of the fluid without outer influence of the field. $S$ is the entropy, $n_i$ is the mole number of each particle, $U_i$ is the partial potential energy of a species i and $\vec{r}$ is the space coordinate. Therefrom, the total differential follows

$$dH^*(S(\vec{r}),\vec{r},n_i(\vec{r})) = \left(\frac{\partial H}{\partial S} + \sum_i \int \frac{\partial U_i}{\partial S} dn_i\right) dS \\ + \left(\frac{1}{A}\frac{\partial H}{\partial \vec{r}} + \frac{1}{A}\sum_i \int \frac{\partial U_i}{\partial \vec{r}} dn_i\right) A d\vec{r} \qquad (2) \\ + \sum_i \left(\frac{\partial H}{\partial n_i} + \frac{\partial U}{\partial n_i}\right) dn_i$$

The derivatives of **(2)** can be identified as

$$T := \frac{\partial H}{\partial S} + \sum_i \int \frac{\partial U_i}{\partial S} dn_i$$

$$-P^* := \frac{1}{A}\frac{\partial H}{\partial r} + \frac{1}{A}\sum_i \int \frac{\partial U_i}{\partial r} dn_i \qquad (3)$$

$$:= -P + \sum_i \int \frac{\partial U_i}{\partial r} \rho_i(r) dr$$

$$\mu_i^* := \frac{\partial H}{\partial n_i} + \frac{\partial U}{\partial n_i} := \mu_i + U_i$$

The definitions are: $T$:=temperature, $\mu_i^*$:=the global chemical potential of a substance (acc. to the original definition of van der Waals and Kohnstam[2]), $P^*$:=global pressure, $\mu_i$:=chemical potential of a substance, $P$:= empirical barometric or hydrostatic pressure, $\rho_i := \partial n_i / A \partial r$:=density of a species of particles and $A$:=unit area.

The phase equilibrium in a potential field can be found if the variation of Hamilton energy $H^*$

$$dH^* = T^g dS^g + T^{li} dS^{li} - P^{g*} dV^g - P^{li*} dV^{li} + \sum_i (\mu_i^{g*} dn_i^g + \mu_i^{li*} dn_i^{li})$$

$$= (T^g - T^{li}) dS - (P^{g*} - P^{li*}) dV + \sum_i (\mu_i^{g*} - \mu_i^{li*}) dn_i = 0 \qquad (4)$$

is minimized to zero in the equilibrium. The second line of the last equation stems from the constraints $dS:=dS^{li} = -dS^g$, $dV:=Adr:=dV^{li} = -dV^g$, $dn_i:=dn_i^{li} = -dn_i^g$ describing the interchange or exchange of entropy, volume and particles between the different phases in adjacent space cells. Therefore, the equilibrium conditions of the extended formalism are

$$T^g = T^{li}$$
$$P^{g*} = P^{li*} \qquad (5)$$
$$\mu_i^{g*} = \mu_i^{li*}$$

Trivially, these equations hold as well in the same phase between adjacent space cells at $r_k$ and $r_{k+1}$. Therefore, the equilibrium conditions above can be rewritten as well in the form

$$\frac{T(r_1)-T(r_2)}{r_1-r_2} = \frac{\partial T}{\partial \vec{r}} = 0$$

$$\frac{P^*(r_1)-P^*(r_2)}{r_1-r_2} = \frac{\partial P^*}{\partial \vec{r}} = 0 \quad\quad (6)$$

$$\frac{\mu_i^*(r_1)-\mu_i^*(r_2)}{r_1-r_2} = \frac{\partial \mu_i^*}{\partial \vec{r}} = 0$$

The phase equilibrium can be interpreted as well as the trivial case result of an optimization acc. to Pontragin's control theory where the total Hamilton energy $H^*(u(\vec{r}))$ is varied using a control variable *u* which is identified as $u := (S(\vec{r}), \vec{V}, n_i(\vec{r}))$ with $\vec{V} := A\, d\vec{r}$. We have to solve

$$\frac{1}{\Delta R_j}\int_0^{\Delta R_j} H^*(u)\, dr_j \rightarrow extremum \quad or \quad \int_0^{\Delta R_j} \delta H^*(u)\, dr_j = 0 \quad\quad (7)$$

The problem can be interpreted as well as a trivial case of a Lagrange variational problem, because $H^*$ does not depend explicitly from other variables than *u* and *u* itself is identified with the "velocity coordinate" of the problem, compare analogous problems in more detail in section 3. Under these very special conditions the Lagrange functional is either stationary either it has an constant optimum value. Therefore, the solution of this problem are the Euler-Lagrange equations

$$\frac{d}{d\vec{r}}\frac{\partial H^*}{\partial S}=0 \;;\quad\quad \frac{d}{d\vec{r}}\frac{\partial H^*}{\partial V}=0 \;;\quad\quad \frac{d}{d\vec{r}}\frac{\partial H^*}{\partial n_i}=0 \quad\quad (8)$$

which are the conditions of equilibrium eq.(**6**) if one remembers the Maxwell relations eq.(**3**). Some of the field extended equations in this section can be found in a different notation in[3].

*Example:*

The state of a real gaseous mixture is set near the critical point. The volume of mixture is rotated at constant velocity in a centrifuge. Due to the centrifugal field, forces appear in the solution which lead to space-dependent profiles of pressure, density and molar ratio. Here a general method is presented how this problem can be solved numerically applying the formalism above.

The equation of state of the fluid at a point $\vec{r}$ is noted here generally by

$$P = P(v, x_i, T; \vec{r}) \tag{9}$$

where $v(\vec{r})$ is the spec. volume, $x_i(\vec{r})$ molar ratio and $\vec{r}$ is the space parameter.

The mixture rule of the mean molecular weight in the centrifugal field $g(r) = \omega^2 r$ is linear, therefore the potential $U$ is

$$U = \int \sum_i n_i M_i g(r) \, dr \tag{10}$$

The total Hamilton energy is

$$H^* = H + U \tag{11}$$

Therefore, the full thermodynamic state of the fluid in every space cell can be characterized by

$$\frac{dH}{d\vec{X}} = \begin{pmatrix} P(\vec{X}) \\ \mu_i(\vec{X}) \end{pmatrix} \qquad i = 1, 2 \ldots n-1 \tag{12}$$

where $\vec{X} = (v, x_i)$ and $T$ is set constant during the calculation and is skipped therefore.

The chemical potentials are

$$\mu_i = \mu_i^0(p^+, T) + RT \ln(f_i/p^+) \tag{13}$$

where $\mu_i^0$ are the standard potentials with $p^+$ as the reference pressure. $f_i$ is the fugacity calculated acc. to $f_i = x_i P \varphi_i$ using a known formula of the fugacity coefficient $\varphi_i$ [4]

$$\ln \varphi_i = Z_M - 1 - \ln Z_M + \frac{1}{RT}\int_V^\infty \left(P - \frac{RT}{v}\right)dv - \frac{1}{RT}\int_V^\infty \sum_{k=1, k \neq i}^{n}\left(\frac{dP}{dx_k}\right)_{T, v, x_j \neq k} x_k dv \tag{14}$$

$Z_M$ is defined as $Z_M = Pv/(RT)$.

For numerical calculation the space of the volume is divided in many infinitesimal small adjacent compartments. If the complete local thermodynamic state (meaning spec. volume v, composition $x_i$, the empirical pressure $P$ and potential $U$) is known in one (reference) compartment k of a vessel in a field it can be concluded on pressure and chemical potential in the adjacent compartment k+1 due to the conditions of phase equilibrium. Due to the equilibrium condition (6) the pressure relation between adjacent space cells at $r_k$ and $r_{k+1}$ is

$$\begin{aligned} P^* &= P(v(r_k), x_i(r_k)) - \int_{r_{ref}}^{r_k} \sum_i \rho_i M_i g(r) dr \\ &= P(v(r_{k+1}), x_i(r_{k+1})) - \int_{r_{ref}}^{r_{k+1}} \sum_i \rho_i M_i g(r) dr \end{aligned} \tag{15}$$

The second equation (15) is in effect the generalized law of hydrostatic or barometric pressure. Analogously for the chemical potentials in adjacent space cells holds

$$\begin{aligned} \mu_i^* &= \mu_i(v(r_k), x_i(r_k)) + \int_{r_{ref}}^{r_k} M_i g(r_k) dr \\ &= \mu_i(v(r_{k+1}), x_i(r_{k+1})) + \int_{r_{ref}}^{r_{k+1}} M_i g(r_{k+1}) dr \end{aligned} \tag{16}$$

In order to obtain the full information of the state in the adjacent space cell k+1, the values of of $(P, u_i)_{k+1}$ have to be calculated from the k$^{th}$ cell acc. to equation **(15)** and **(16)**.

$$\begin{pmatrix} P \\ \mu_i \end{pmatrix}_{k+1} = \begin{pmatrix} P(\vec{X}_{k+1}) \\ \mu_i(\vec{X}_{k+1}) \end{pmatrix} \qquad i=1,2 \ldots n-1 \tag{17}$$

Then **(17)** has to be solved for $\vec{X}_{k+1}$. For the purposes of numerical calculation (in order to avoid inaccuracies due to the integral in the calculation of $P^*$), however, it is recommended to take an other equivalent representation of the thermodynamic state. Due to the phase equilibrium conditions (eq.**(6)**) and eq.**(3)** it holds

$$\frac{\partial}{\partial x_i} \frac{1}{\rho} \frac{\partial P}{\partial r} = M_i g(r) = \frac{\partial \mu_i}{\partial r} = \frac{\partial}{\partial x_i} \frac{\partial G}{\partial r} = \frac{\partial}{\partial x_i} \left( \frac{1}{\rho} \frac{\partial P}{\partial r} - S \frac{\partial T}{\partial r} \right) \tag{18}$$

which proves 1) $T$ =constant and 2) that the equation for $P$ and the sum $\Sigma \mu_i x_i$ is linear dependent. Therefore, the equation for $P$ in eq. **(17)** can be replaced by the equation for $\mu_n$ and instead of eq. **(17)** the following system of equations has to be solved

$$\left( \mu_i \right)_{k+1} = \left( \mu_i(\vec{X}_{k+1}) \right) \qquad i=1,2 \ldots n \tag{19}$$

If the equation of the thermodynamic state is solved for $\vec{X}$ we have the full information about the fluid in the adjacent space compartment.

This iteration procedure is repeated over all space cells until the thermodynamic state of the whole volume in the cylinder is determined completely.

Sometimes, under practical conditions no initial or reference values of the thermodynamic state is known in any compartment. However, the total mass and the total molar ratios are known. Then, additional equations describing the mass conservation of the different particles

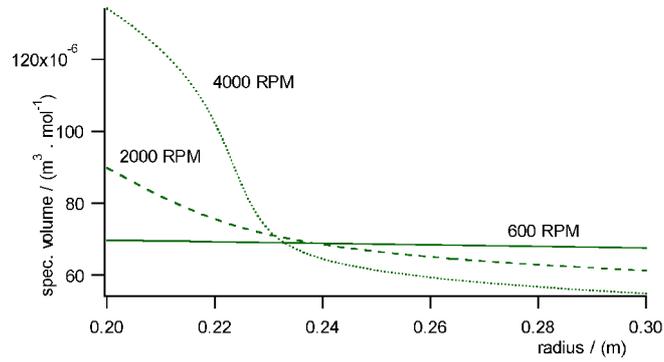

Fig.1a: Spec. volume profile in a rotating vessel versus radius r at different rotation speeds
Initial state without field: Argon - Methane 55 bar, molar ratio $x_1$ (Argon)= 0.56, temperature 170 K, vessel: inner edge $r_1$ =20cm and outer edge $r_2$ =30cm

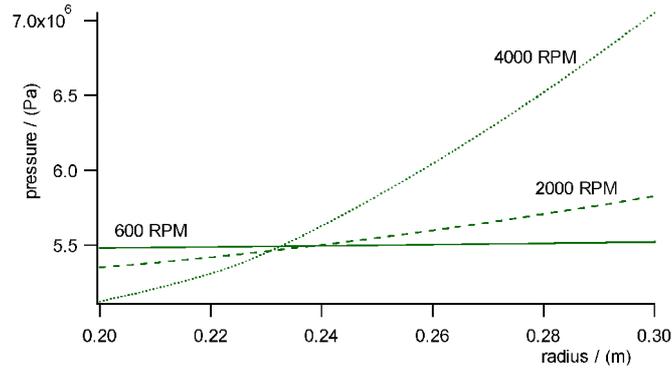

Fig.1b: Pressure profile versus radius in a rotating vessel at different rotation speeds
Initial state without field: Argon - Methane 55 bar, molar ratio $x_1$ (Argon)= 0.56, temperature 170 K, vessel: inner edge $r_1$ =20cm and outer edge $r_2$ =30cm

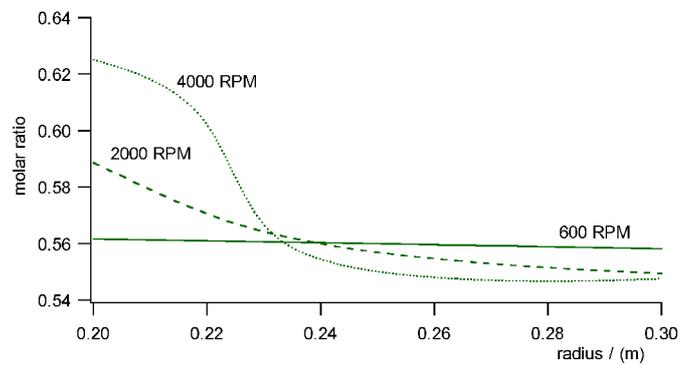

Fig.1c: Molar ratio profile $x_1$ versus radius at different rotation speed. in a rotating vessel
Initial state without field: Argon - Methane 55 bar, molar ratio $x_1$ (Argon)= 0.56, temperature 170 K vessel: inner edge $r_1$ =20cm and outer edge $r_2$ =30cm help to determine the full thermodynamic state which is shown in more detail in appendix 1.
As an example numerically calculated profiles of spec. volume $v$, molar ratio $x_i$ and pressure $P$ near the critical point are shown for the system Argon-Methane in fig. 1a)-c).

## 3. The statistical derivation of the thermodynamic formalism with fields

It is well known that the mechanic equations of motion can be found as solutions of a Lagrange variational problem. The solution is obtained, if the functional L has an extremum, where $x(t_0)=x_0$ and $x(t_1)=x_1$ are start and end point of the path.

$$I(x) = \int_{t_1}^{t_0} L(x,\dot{x};t)\, dt = Extremum \tag{20}$$

The variational problem of mechanics can be regarded as well as a special case of a general problem of control theory [5] where the control variable $u(t)$ coincides with the velocity variable $\dot{x} = f(u,x;t) = u$. The functional of this special control theory problem is

$$J(x) = \int_{t_1}^{t_0} L(x,u;t)\, dt = Extremum \tag{21}$$

with $\dot{x}=u$, $x(t_0)=x_0$, and $x(t_1)=x_1$. The Hamiltonian is defined to

$$H = pu - L(x,u;t) \tag{22}$$

The adjunct system is defined to

$$\dot{p} = L_x(x,u;t) \tag{23}$$

Acc. to control theory the problem can be solved if the extremum of the Hamiltonian is found

$$H_u = p - L_u(x, u; t) = 0 \qquad (24)$$

If this equation is differentiated for t the solution is the Euler-Lagrange equation

$$\frac{d}{dt} L_{\dot{x}} - \frac{dL}{dx} = 0 \qquad (25)$$

because of $\dot{x} = u$ and $\dot{p} = L_x$ as defined above.

The Hamiltonian has here a maximum for the chosen coordinates because $H_{uu} < 0$.

In ref. [6] proofs of the different versions of Pontragin's maximum principle can be found.

Similarly, as shown by Landau and Lifshitz [1], the Lagrangian of electrostatics can be varied with respect to the electric coordinates **E** or **D**. If only one Maxwell relation is given the other can be reconstructed by the variational formalism applied to any thermodynamic potential. These results could be embedded in a more general mathematical framework [7] which derives general relativity including all sub-theories using a Lagrange energy approach developed to second order.

Therefore, because a thermodynamic system is a mechanic many particle system in a field, similar variational features of the thermodynamic Hamiltonian should be expected.

In equilibrium, due to energy conservation, the Hamiltonian is constant. Then, the mechanical time mean of the Hamiltonian is trivially identical to the Hamiltonian as well.

Acc. to the statistical approach the time mean of total energy of all particles is identical to the ensemble average. Or in mathematic language (using the symbol $\Delta T$ as measuring time interval)

$$H_{tot} := \frac{1}{\Delta T} \int_0^{\Delta T} \sum_i (\varepsilon_i + U_i + \sum_{j \neq i} U_{ij}) dt$$
$$= \iint W(\varepsilon_K, n_i) \, \varepsilon_K \, d\varepsilon_K dn_i \quad (26)$$
$$= \iiint \hat{W}(\varepsilon, n_i(\vec{r}), \vec{r}) \, \varepsilon \, \rho(n_i(\vec{r}), \vec{r}) \, dV(\vec{r}) d\varepsilon \, dn_i(\vec{r})$$

$W(\varepsilon_K, n_i)$ is here the canonic probability distribution in the whole volume as defined in Mayers book [8] p.6+7 with the eigenvalue of the total Hamilton energy $\varepsilon_K$.

In the third line $\hat{W}(\varepsilon, n_i(\vec{r}), \vec{r})$ is the probability function in a volume element $dV(\vec{r})$ at the point $\vec{r}$ and $\rho(n(\vec{r}), \vec{r})$ is a weighting function in which characterizes the total particle number density profile. The mean Hamilton energy $\varepsilon$ at a point in the field is

$$\varepsilon := \varepsilon_{kin} + U_{int}(n_i(\vec{r}), \vec{r}) + U(n_i(\vec{r}), \vec{r}) \quad (27)$$

where $\varepsilon_{kin}$ is the kinetic energy in the field, $U_{int}$ is the mean field potential between the particles due to the real fluid behaviour and U is the energy due to the field from outside. As shown above by control theory the mechanic Hamiltonian has an extremum. If this feature is transferred from mechanics to thermodynamic notation as an ansatz analogously, then at every space cell the Hamilton energy density function

$$\hat{\mathcal{H}} := \hat{W}\hat{H} := \hat{W}\varepsilon\rho \quad (28)$$

should be a extremum as well.

For the following calculation all symbols are made dimensionless by defining

$$\mathcal{H}' := \hat{\mathcal{H}} V_0 kT/\mu^* \quad W' := -\hat{W}(kT)^2/\mu^* \quad V_0 := kT/P^* \quad \rho' := \rho V_0$$
$$\varepsilon' := \varepsilon/kT \quad n_i' := -\mu^* n_i/kT \quad V' := V/V_0 \quad (29)$$

Now, the calculation from **(20)** -> **(25)** can be done analogously in thermodynamics as well if the mechanical variables are exchanged by thermodynamical terms acc. to the following table eq. **(30)** below (using the definition $\vec{1} := (1,1,\ldots,1)$):

$$H := \sum_i \left( \frac{p_i^2}{2m_i} + \sum_{j \neq i} V(x_i - x_j) \right) \quad \rightarrow \quad \mathcal{H}' := \rho' \varepsilon' W' := E^* W'$$

$$x_i \quad \rightarrow \quad E^* := \rho' \varepsilon'$$

$$t \quad \rightarrow \quad \vec{t} := (\varepsilon', n'_i, V') \quad (30)$$

$$\dot{x}_i = f(x_i, u_i) = u_i \quad \rightarrow \quad \vec{u} := \left( \frac{\partial E^*}{\partial \varepsilon'}, \frac{\partial E^*}{\partial n'_i}, \frac{\partial E^*}{\partial V'} \right)$$

$$L = p_i u_i - H \quad \rightarrow \quad \vec{L}'_{\vec{t}} := p_{\vec{t}} \cdot \vec{u} - \mathcal{H}' \cdot \vec{1}$$

The reduced Hamiltonian $\mathcal{H}'$ is optimized for every variable of $\vec{t}$ separately

$$H_{total} \sim \begin{pmatrix} \int W' \rho' \varepsilon' \, d\varepsilon' \\ \int W' \rho' \varepsilon' \, dV' \\ \int W' \rho' \varepsilon' \, dn'_i \end{pmatrix} \rightarrow maximum \quad (31)$$

The adjunct variables $p_{\vec{t}}$ are defined by the equations

$$\frac{\partial p}{\partial \varepsilon'} = -W'$$

$$\frac{\partial p}{\partial V'} = -W' \quad (32)$$

$$\frac{\partial p}{\partial n'_i} = -W'$$

The Lagrangians $\vec{L}$ of this problem follow from the last line of eq.**(30)**

$$L'(\varepsilon') = W' \frac{\partial E^*}{\partial \varepsilon'} - W' E^*$$

$$L'(V') = W' \frac{\partial E^*}{\partial V'} - W' E^* \quad (33)$$

$$L'(n'_i) = W' \frac{\partial E^*}{\partial n'_i} - W' E^*$$

Therefrom, three separate differential equation are obtained

$$\left(\frac{\partial}{\partial \varepsilon'}\frac{\partial}{\partial E^*_{\varepsilon'}} - \frac{\partial}{\partial E^*}\right)(L'_{\varepsilon'}) = \frac{\partial W'}{\partial \varepsilon'} + W' = 0$$

$$\left(\frac{\partial}{\partial V'}\frac{\partial}{\partial E^*_{V'}} - \frac{\partial}{\partial E^*}\right)(L'_{V'}) = \frac{\partial W'}{\partial V'} + W' = 0 \quad (34)$$

$$\left(\frac{\partial}{\partial n'_i}\frac{\partial}{\partial E^*_{n'_i}} - \frac{\partial}{\partial E^*}\right)(L'_{n'_i}) = \frac{\partial W'}{\partial n'_i} + W' = 0$$

The combined solution of these differential equations **(34)** is the distribution function

$$W' = W'_0 \exp[-(\varepsilon' + V' + \sum n'_i)] \quad (35)$$

If the definitions of the reduced variables **(29)** are reinserted one obtains

$$W'(n_i, \varepsilon) = W'_0 \exp[(-\varepsilon - P^*V + \sum \mu^*_i n_i)/kT] \quad (36)$$

This is Mayer's master equation [8] which is extended here for systems containing space-dependent potential fields. The norm is chosen to be

$$\int\int W'(n', \varepsilon')dn'd\varepsilon' = \int\int \hat{W}(n, \varepsilon)dn d\varepsilon = 1 \quad (37)$$

Therefrom, using **(29),** the standard representation of thermodynamics can be derived acc. to the procedure presented in Mayer's book [8] p.8. The mean number of particles in a volume is then

$$\bar{n}_i = \int\int n_i \hat{W}(n_i, \varepsilon)dn_i d\varepsilon \quad (38)$$

The mean (always indicated by a bar) of Gibbs's free energy $\bar{G}$ is

$$\bar{G} = \iint \sum \hat{W}(n_i, \varepsilon) n_i \mu_i^* dn_i d\varepsilon \qquad (39)$$

From the master equation **(36)** and **(29)** is derived by differentiation

$$\frac{\partial \hat{W}}{\partial \mu_i^*} = (kT)^{-1} [n_i - V(\partial P^*/\partial \mu_i^*)] \hat{W}$$

$$T \frac{\partial \hat{W}}{\partial T} = (kT)^{-1} [P^* V - \sum n_i \mu_i^* + \varepsilon - VT(\partial P^*/\partial T)] \hat{W} \qquad (40)$$

Both equations are summed up over all possibilities $\hat{W}$ which give in sum 1 acc. to eq. **(37)**. Therefore, each sum of all derivatives is zero and the relations between the corresponding mean values (indicated by a bar) are

$$\frac{\partial \bar{P}^*}{\partial \mu_i} = \bar{n}_i / V$$

$$\frac{\partial \bar{P}^*}{\partial T} = (VT)^{-1} [\bar{H}^* + \bar{P}^* V - \bar{G}^*] = \frac{\bar{S}}{V} \qquad (41)$$

From the second equation **(41)** follows Shannon's definition of entropy

$$\bar{S} = k \int W' \ln(W') dn_i' d\varepsilon' \qquad (42)$$

It should be noted that the entropy of the field extended formalism of thermodynamics has the same expression as without field. Therefrom, it can be concluded that the entropy of the field extended thermodynamic formalism follows the maximum entropy principle as well, because the proofs in textbooks like [9] can be applied accordingly.

## 4. Second law violations due to potentials with saddle points

The minimum principle of potentials is derived in many textbooks of thermodynamics from the second law [4]. In this section this procedure is reversed and the second law is derived from the extremum behaviour of the Hamilton energy. As shown in [1] the second derivative of the Hamilton energy can be obtained from the second derivative of the mathematical variational or control problem and gives information about the direction of the irreversibilities in a potential field. This information comes from mathematics alone and is independent from any additional empirical or axiomatic input information like second law. Therefore, the question has to be discussed whether the mathematical and the axiomatic approach of equilibrium thermodynamics are equivalent.

Both approaches make the same prediction for thermodynamic standard cases where H is convex and irreversibilities obey always $dH<0$. However, if saddle points exist, H can obey $dH>0$ in certain directions of the state space. Then, it will be shown in the following that Clausius's version of second law can be "reversed".

The energies H' and H" of capacitively loaded thermodynamic systems are defined [1] by

$$\begin{aligned}
H'(S,V,n_i,\mathbf{P}) &= H(S,V,n_i) + \iint \mathbf{E}\, d\mathbf{P}\, dV \\
&= H(S,V,n_i) + \frac{1}{2}\int \frac{\mathbf{P}^2}{\varepsilon_0(\varepsilon-1)}\, dV \\
H''(S,V,n_i,\mathbf{E}) &= H(S,V,n_i) - \iint \mathbf{P}\, d\mathbf{E}\, dV \\
&= H(S,V,n_i) - \frac{1}{2}\int \varepsilon_0(\varepsilon-1)\mathbf{E}^2\, dV
\end{aligned} \quad (43)$$

where the definitions are $\varepsilon_0(\varepsilon-1) :=$ dielectric constant, $\mathbf{E} :=$ electric field and $\mathbf{P} :=$ electric polarisation. The same formulas can be written in differentials

$$dH´(S,V,n_i,\mathbf{P})=dH(S,V,n_i)+\int(\mathbf{E}d\mathbf{P})dV$$
$$dH´´(S,V,n_i,\mathbf{E})=dH(S,V,n_i)-\int(\mathbf{P}d\mathbf{E})dV \quad (44)$$

Regarding the 2nd derivative of *H'* and *H''* with respect to the electric variables **P** or **E**, comp. eq.(**43**), both of these potentials approach an thermodynamic extremum for irreversible processes into thermodynamic equilibrium. For constant homogeneous dielectrics the potential $H'(S,V,n_i,\mathbf{P})$ has a minimum, and $H''(S,V,n_i,\mathbf{E})$ has a maximum, if $\varepsilon_0(\varepsilon-1)>0$, $dn_i=0$, $dS=0$ and $dV=0$. Therefore, acc.to [1], the following unequalities hold for irreversible changes of state

$$\Delta H´_{irrev}(\mathbf{P})<0 \quad for\ S,V,n_i=constant$$
$$\Delta H´´_{irrev}(\mathbf{E})>0 \quad for\ S,V,n_i=constant \quad (45)$$

Due to the Legendre transformation formalism analogous expressions of eq.(**43**) and (**44**) hold for free enthalpy, i.e.

$$dG´(P,T,n_i,\mathbf{P})=dG(P,T,n_i)+\int(\mathbf{E}d\mathbf{P})dV$$
$$dG´´(P,T,n_i,\mathbf{E})=dG(P,T,n_i)-\int(\mathbf{P}d\mathbf{E})dV \quad (46)$$

and

$$\Delta G´_{irrev}(\mathbf{P})<0 \quad for\ T,P,n_i=constant$$
$$\Delta G´´_{irrev}(\mathbf{E})>0 \quad for\ T,P,n_i=constant \quad (47)$$

In words, the equations (**45**) and (**47**) can be interpreted as a "electric Chatelier-Braun-principle", which states for simple dielectrics that they tend to discharge themselves. Because (**47**) will lead to second law violations it should be emphasized that the correctness of both equations **(47)** is based either theoretically on the variation principle either experimentally on material data as shown fig.2a and in section 4.

Now, an electric cycle is regarded which has an irreversible path into a maximum of $G''$ at constant field **E**, comp. fig.2b and fig.3:

Because G'' is a potential for any closed cycle over three points(1->2->3->1) one obtains

$$\Delta G'' = \Delta^{312}G''_{rev} + \Delta^{23}G''_{irrev} = 0 \qquad (48)$$

According to the extremum principle(eq.**(47)**) $\Delta^{23}G''_{irrev} > 0$ holds. Due to **(48)** follows $\Delta^{312}G''_{rev} < 0$. Because of the isofield ($\mathbf{E}_2 = \mathbf{E}_3$) irreversible change of state (2->3) it holds also $\int_{E_3}^{E_2} \mathbf{P}\, d\mathbf{E} = 0$. This zero expression is added to the second formula of **(46)** which yields

$$\Delta^{312}G''_{rev} = \int - (\int_{3\to 1}^{2} \mathbf{P}\, d\mathbf{E} + \int_{2}^{3} \mathbf{P}\, d\mathbf{E})\, dV = \int - (\oint \mathbf{P}\, d\mathbf{E})\, dV = \int (\oint \mathbf{E}\, d\mathbf{P})\, dV < 0 \qquad (49)$$

The sign of the right integral shows indicates a "gain" cycle which is reversed compared to the hysteresis of a ferroelectric substance. Because the cycle proceeds isothermically (with only one heat reservoir) the Clausius statement of second law is violated.

The proof is as follows: Due to energy conservation and because H'' is a potential

$$\oint H'' = \int (-\oint \mathbf{P}\, d\mathbf{E})\, dV + \oint T\, dS = 0 \qquad (50)$$

holds. Therefrom, because the electric term is negative, comp. eq.**(49)**, it follows that the net heat exchange $\oint T\, dS$ has a positive sign. This implies $\oint dS > 0$ because $T = constant$. This means that the cycle takes heat from the environment and gives off electrical work under isothermal conditions which is contrary to the Clausius formulation of the 2nd law. □

It should be noted that this proof is not in contradiction to the principle of maximum entropy. Acc. to the calculation holds

$$\oint S = \Delta S_{3\to 1\to 2}^{rev} + \Delta S_{2\to 3}^{irrev} > 0 \tag{51}$$

Due to the maximum entropy principle it can be said that the entropy is higher at state 3 than in the state 2. Therefore, it holds

$$\Delta S_{3\to 1\to 2}^{rev} < 0 \tag{52}$$

meaning that heat is given off on the path 3->1->2 to the system environment which can be calculated here from the thermodynamic potential because the entropy is a clearly defined function of state on the reversible path. For the irreversible path, however, the entropy difference $\Delta S_{2\to 3}^{irrev}$ must be calculated by applying energy conservation **(50)**. Therefore, the following inequalities hold due to **(51)** and **(52)**

$$\Delta S_{2\to 3}^{irrev} > -\Delta S_{3\to 1\to 2}^{rev} > 0 \tag{53}$$

This example illustrates that different formulations of second law may be not equivalent, comp.[10].

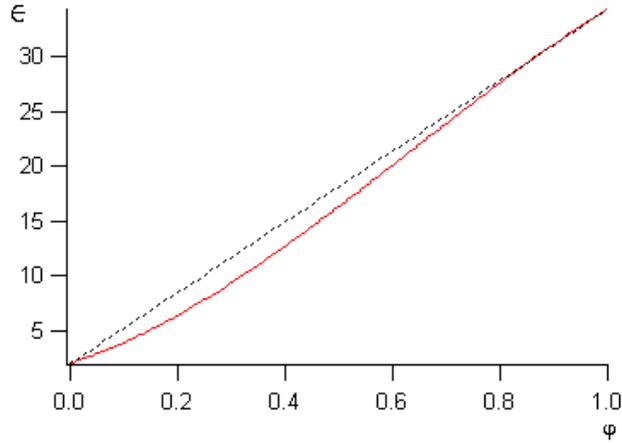

Fig.2a: Dielectric constant $\epsilon$ vs. volume fraction $\varphi$ of the mixture nitrobenzene- 2,2,4-nitropenthane at 29.5 C from [11]. The curve follows the series $\varepsilon = 2.1 + 15.1\varphi + 36.5\varphi^2 - 19.4\varphi^3$.
Because of $\partial^2\varepsilon/\partial\varphi^2 > 0$ mixture processes in strong electric homogeneous field can obey $\Delta G''(\mathbf{E}) = -\varepsilon_0 \Delta\varepsilon(\varphi) \mathbf{E}^2 V/2 > 0$ for $T, P, \mathbf{E}, n_i$ = constant. and $\varepsilon$ linear with resp. to $\mathbf{E}$,
Note that this is a material property which can violate Clausius's version of second law!     comp. text and as well fig.2b), fig. 3 and fig.4d).

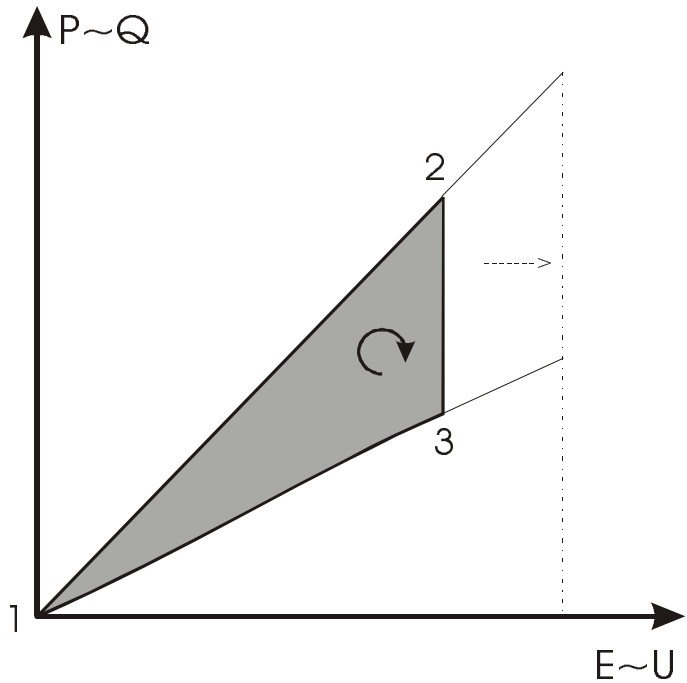

Fig.2b: Qualitative diagram of polarisation **P** vs. electric field **E** or charge Q vs. voltage U
Isotherm electric cycle of a material with $\partial^2\varepsilon/\partial\varphi^2 > 0$.
1->2 charging the capacitance with voltage;     2->3 mixing both components by opening a tap, comp. fig 3.     3->1 discharging the capacitance
Acc. to the diagram the orientation of the working area shows a gain hysteresis. The energy output can be enhanced by increasing **E** or U. Therefore, for very strong fields this output can be higher than the constant energy $\Delta E$ input necessary to separate the components at zero field by centrifugation or chemical separation.     comp. fig.2a) and 3

## 5. *About the concrete realization of second law violating cycles*

The following example shows principally the possible existence of a second law violating cycle, comp. fig.3:

Step 1: The cycle starts -for example- with a 50:50 mixture of two liquid or gaseous substances. It is assumed that the dielectric constant of the mixture has a mixture rule with $\partial^2 \varepsilon / \partial \varphi^2 > 0$ ($\varphi$:=volume fraction of one component). The mixture will be separated (for instance by centrifugation) into two halves of 40% and 60% of concentration. Therefore, a chemical or mechanical input energy $\Delta E$ is necessary for the separation.

Step 2: Then, both halves of different concentration are used as dielectrics and are loaded parallel with the same strong electric field **E**.

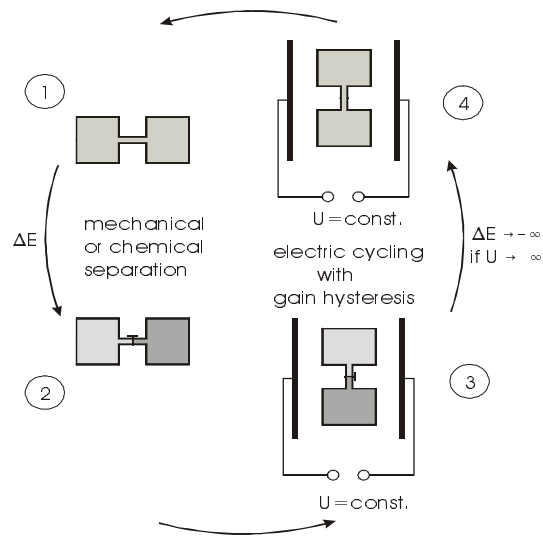

Fig.3: Principle of a second law violating cycle with a nonlinear dielectrics, comp. fig.2a .
1) starting point, mixture 50:50 -> 2) after separation procedure 40:60 after energy input ΔE for instance by centrifugation -> 3) after applying a strong electric field -> 4) after the mixing process in the field -> 1) after discharging the capacitance. The work area of the capacitance, comp. fig 2b), can principally overcome the energy of the separation if the field is high enough.

Step 3: After the charging process both halves are remixed in the field. Due to this procedure the total dielectric constant decreases due to the material property $\partial^2\varepsilon/\partial\varphi^2 > 0$. Therefore, a current flows from the capacitance at constant voltage.

Step 4: Then, the whole capacitance is discharged and the cycle starts again after the necessary relaxation time.

As shown in fig. 2b the electric work diagram of the capacitance shows a "gain" hysteresis. In principle, in this simple model, the work gain area can be driven to infinity if the electric field is driven to infinity. Therefore, the work gain of the electric cycle can be higher than the defined energy input $\Delta E$ for separation. Due to energy conservation the differing amount of energy between input and output has to come from the heat of environment. □

Of course the realisation of this cycle is not so easy as the first idea but there exist some experimental data which allow to get a first feeling of about what seems to be possible.

In 1965 Debye and Kleboth [11] investigated the influence of electric fields on phase equilibria of liquids. They observed the following facts :

1) The influence of homogeneous fields on phase equilibria is weak due to the big difference between the electric field energies applied compared with the chemical energies involved in the mixture process. Comparing electric field energies with thermal energies

$$\frac{1}{2}\varepsilon_0(\varepsilon-1)\mathbf{E}^2 v = RT \qquad (54)$$

one obtains $\mathbf{E} = 6{,}13 \cdot 10^8$ V/m if values for water at room temperature are inserted in **(54)**. This is higher than the breakdown voltage of stronger bulk plastic isolator materials like Polyoxymethylen or Polyethylenenterphthalat [12] which can resist to more than $4{,}5 \cdot 10^7$ V/m. Therefore, in order to avoid this principal problem it is recommended to look for effects in the neighbourhood of a critical point.

2) Debye and Kleboth found mixtures whose phase diagram was influenced by strong homogeneous fields. In order to obtain a field-induced decrease of the critical temperature in a phase diagram of a solution, a nonlinear mixture rule of the dielectric constant with $\partial^2\varepsilon/\partial\varphi^2 > 0$ was necessary, comp.fig.2a).

Debye and Kleboth investigated the turbidity of a solution nitrobenzene - 2,2,4-trimethylpentane at critical concentration and temperatures slightly above the critical temperature. If a strong field was applied to the solution the turbidity decreased confirming that the critical temperature of the phase diagram of the mixture was shifted by the field to lower values which was predicted by their theoretical considerations as well.

Similar investigation in homogeneous fields (but below the critical point) were done with polymer solutions by Wirtz and Fuller [13] [14]. They investigated electrically induced sol-gel phase transitions. To explain their experiments they used a Flory-Huggins model [13] extended by an electric interaction term. They find that this model describes the qualitative behaviour of their solutions correctly.

It can be shown for these solutions that an second law violating isothermal cycle is possible, which is the electric analog of a Serogodsky or a van Platen cycle of binary mixtures discussed recently in [15]. Compared with the example above, comp. fig.2a)-b) and fig.3, the separation of the components is achieved here by the electrically induced phase transition due to the nonlinearities of the dielectrics. No other thermodynamic, chemical or mechanic processes are necessary to obtain the separation of the components of the mixture.

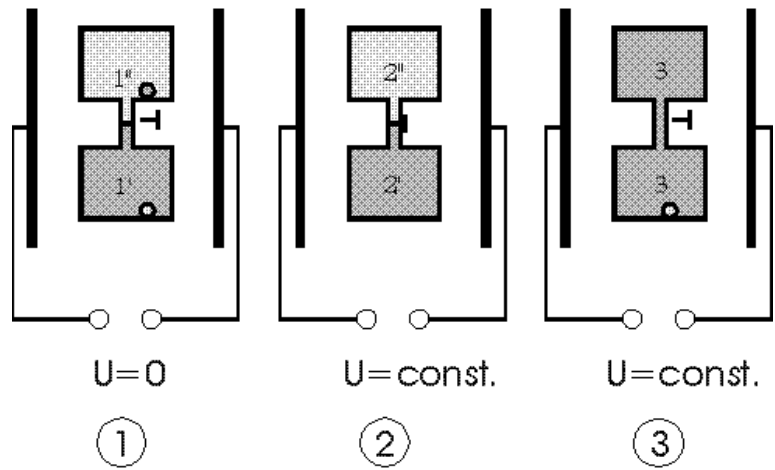

Fig 4a: Isothermic isobaric electric cycle of a diluted polymer solution as dielectric
  1) voltage U=0: system in 2-phase region    2) both volumes separated, rise
  of voltage from zero to U=const.: each volume compartment in 1-phase region
  3 ) voltage U=const.: opening the tap and returning to the phase separation line
  by remixing

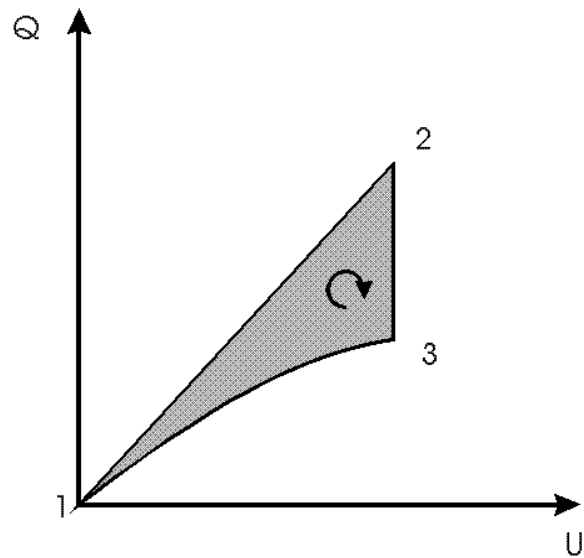

Fig.4b: Isothermal electric cycle in capacitor due to electrically induced phase transitions;
  charge Q vs. voltage U plotted;
  1 starting at 2- phase region line with zero field, 1-2 applying a field
  with tap closed,  2 opening the tap, 2-3 discharging and remixing in field,
  3 returning to starting point 1 by discharging the capacitor; a gain
  work area is predicted due to dG"(**E**)>0 during the irreversible mixing process.

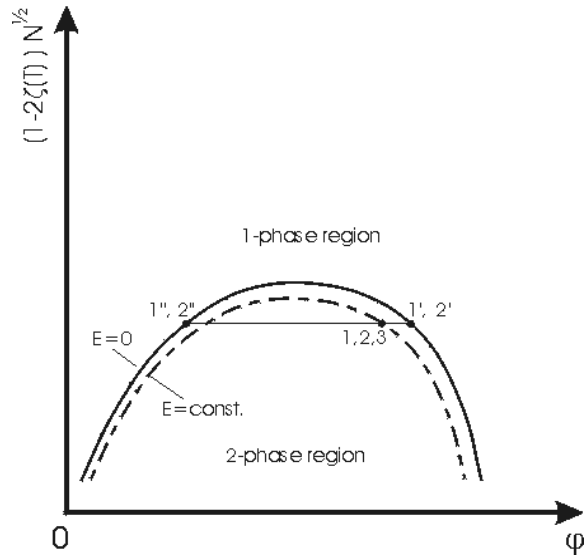

Fig.4c: $[1-2\zeta(T)]N^{0.5}$ vs. volume fraction φ (with $T$ :=temperature)
phase diagram of a polymer solution with and without electric field **E**
according to [13,14]; plot shows a modified Flory-parameter versus volume fraction φ
of polymers; points 1: **E**=0, 2-phases, both points 1 at the phase
separation line; points 2: **E**=const., both points 2 of the splitted volume
in 1- phase region; point 3: **E**=const., after opening the tap: point 3 is on
the phase separation line

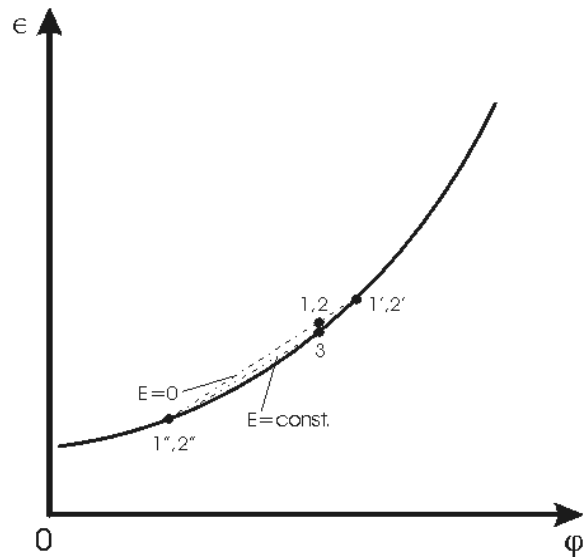

Fig.4d: Dielectric constant ε vs. volume fraction φ of polymers in a dilute solution;
points 1, 2 and 3 refer to points in fig.4a)-c). According to the theory [13,14]
$d^2\epsilon/d\varphi^2 > 0$ holds near the critical point. Therefore ε(φ) has to turn to the left and
the dielectric constant has to decline during remixing 2->3. Observations at the similar
system[11] support this prediction, comp . Fig.2a)

The system is the experimental realization of the model system in the proof in section 4. This closed splitted cycle is proceeded with a capacitor using as dielectrics a sol-gel mixture like polystyrene in cyclohexane (upper critical point solution) or p-chlorostyrene in ethylcarbitol (lower critical point solution). The composition of the solution is separated periodically by a demixing phase transitions induced by switching off the field. After the separation of both phases by splitting into two volumes they are remixed again (irreversible path during the cycle !) after opening the separating tap in a strong field.

The cycle is started in the 2- phase region at zero field at the points 1, comp.fig. 4a)-c), where the volume is split by closing the tap separating both phases. Then a strong homogeneous electric field **E** is applied. At the point 2 the solution is separated in two phases each in a different compartment. This is represented by the points 2' and 2'' in the fig. 4. Then, the tap is opened and the solutions of both compartments are mixed. During the mixing (2->3) the electric field is kept constant by discharging the capacitor during the decline of the dielectric constant, cf. fig.4d). Then, in the phase diagram fig.4c) and as well in fig. 4a) +b), the mixed solution is at the phase separation line at point 3. In the last step of the cycle the capacitor is discharged completely and the system goes back into the 2-phase region to point 1 and demixes. According to the theory, eq.**(47)**,

$$\Delta^{23}G''_{irrev} = -\Delta^{312}G''_{rev} = G''_3 - G''_2 > 0 \qquad (55)$$

Now, $S^* := V'/V$ is defined as the splitting factor of the total volume $V$. $V'$ and $V''$ are the volumes of the compartments each where $V := V' + V''$. The difference of the free enthalpy is written using the definition or $G''$ in **(46)** assuming $\varepsilon_0(\varepsilon-1)$ to be dependent from $\varphi$ and independent from **E**

$$\Delta^{312}G''_{rev} = -\frac{1}{2}\mathbf{E}_2^2 \varepsilon_0 [\varepsilon(\varphi_{(1'\to 2')}).S^* + \varepsilon(\varphi_{(1''\to 2'')}).(1-S^*)]V$$
$$+ \int_0^{\mathbf{E}} V \varepsilon_0 [\varepsilon(\varphi_{(1'\to 3')}).S^* + \varepsilon(\varphi_{(1''\to 3'')}).(1-S^*)]\mathbf{E}d\mathbf{E} \qquad (56)$$

The right side of the first line represents the stored linear combined field energy 1->2 of the separated volume parts (points 2' and 2") at point 2, the second line stands for the field energy difference (1->3) of both the connected compartments containing the coexisting phases φ' and φ". In the first line $S^*$, ε, φ' and φ" are constant, in the second line $S^*$, ε, φ' and φ" are dependent from **E** in the 2-phase region.

Wirtz et al. [13,14] apply a Flory free-energy density approach of an incompressible dilute monodisperse polymer solution to describe their systems. The "ansatz" is here

$$f'(\mathbf{P}) = \frac{1}{v_m}\left((\varphi/N)\ln\varphi + \frac{1}{2}(1-2\zeta(T))\varphi^2 + \frac{1}{6}\varphi^3\right) + \frac{(\beta \mathbf{P}^2/2)}{\varepsilon_0(\varepsilon(\varphi)-1)}$$
$$f''(\mathbf{E}) = \frac{1}{v_m}\left((\varphi/N)\ln\varphi + \frac{1}{2}(1-2\zeta(T))\varphi^2 + \frac{1}{6}\varphi^3\right) - (\beta \mathbf{E}^2/2)\varepsilon_0(\varepsilon(\varphi)-1) \qquad (57)$$

where $N$:= polymerisation number, $\varepsilon_0$:= dielectric constant of vacuum, ε := dielectric constant of the material, β := $1/(kT)$ with $k$ :=Boltzmann number, $v_m$ :=monomer volume and ζ(T) :=Flory parameter which depends on the temperature T and the material.

Due to $W_{el} = \int UI dt = \int U dQ = \iint \mathbf{E}d\mathbf{P}dV = 1/2.\int \mathbf{P}^2/(\varepsilon_0(\varepsilon-1))dV$ the electrical field term of the first formula **(57)** is proportional to the electrical work $W_{el}$ applied to the system. In the second formula **(57)** the electrical field term is proportional to the negative field energy which is the potential perceived by the dipolar matter. This equation is equivalent to the Hamilton energy eq.**(1)**. The chemical potential per volume $\mu^*$ (**E**) follows from **(57)**

$$\mu^*(\mathbf{E}) := \left.\frac{\partial f^{//}}{\partial \varphi}\right|_{\mathbf{E}} = \mu^+(\mathbf{P}) := \left.\frac{\partial f^{/}}{\partial \varphi}\right|_{\mathbf{P}}$$
$$= \frac{1}{v_m}\left(\frac{1}{N} + \frac{1}{N}\ln\varphi + (1-2\zeta)\varphi + \frac{\varphi^2}{2}\right) - \frac{\beta\varepsilon_0\mathbf{E}^2}{2}\frac{\partial\varepsilon}{\partial\varphi} \quad (58)$$

However, both these formulas are not appropriate to calculate stability condition and phase diagram because a derivation with **(58)**, comp. eq. **(60)** and **(61)**, yield wrong results if one compares it with the experiment and the calculation of [11] and [13].

Therefore, [11] and [13] choose another dependence of the chemical potential on the electric field. With the abbreviation $f_0 := ((\varphi/N)\ln\varphi + (1-2\zeta)\varphi^2/2 + \varphi^3/6)/v_m$ and the definitions $f^{/}(\mathbf{E}) := f_0 + 1/2 \cdot \varepsilon_0(\varepsilon-1)\mathbf{E}^2$ the relevant chemical potential per volume is

$$\mu^+(\mathbf{E}) := \partial f^{/}(\mathbf{E})/\partial\varphi = \frac{1}{v_m}\left(\frac{1}{N} + \frac{1}{N}\ln\varphi + (1-2\zeta)\varphi + \frac{\varphi^2}{2}\right) + \frac{\beta\varepsilon_0\mathbf{E}^2}{2}\frac{\partial\varepsilon}{\partial\varphi} \quad (59)$$

The stability condition in a constant field is accordingly

$$\partial^2 f^{/}(\mathbf{E})/\partial\varphi^2 = \partial\mu^+(\mathbf{E})/\partial\varphi > 0 \quad (60)$$

A critical point is defined by the equations

$$\begin{array}{l}\partial^2 f^{/}(\mathbf{E})/\partial\varphi^2 = \partial\mu^+(\mathbf{E})/\partial\varphi = 0 \\ \partial^3 f^{/}(\mathbf{E})/\partial\varphi^3 = \partial^2\mu^+(\mathbf{E})/\partial\varphi^2 = 0\end{array} \quad (61)$$

Both definitions are supported by calculation and experiments of [11] and [13].

Consequently, the phase equilibrium is determined by the equations [13]

$$\mu_0^+ = \mu^+(\mathbf{E})(\varphi^{/}) = \mu^+(\mathbf{E})(\varphi^{//})$$
$$\int_{\varphi^{/}}^{\varphi^{//}} [\mu_0^+ - \mu^+(\mathbf{E})(\varphi)]d\varphi = \int_{\varphi^{/}}^{\varphi^{//}} \left.\frac{\partial\mu^+}{\partial\varphi}\right|_{\mathbf{E}} \varphi\, d\varphi = 0 \quad (62)$$

The first equation describes the chemical potential to be equal in both phases. The second equation is the Maxwell construction applied to the chemical potential $\mu^+$. The solution of this system of equations can be done numerically. Qualitative results are in fig.4c).

(It should be noted, that the ansatz **(57)** and **(59)** above differs from the references [11] and [13] : First, both authors use an excess energy term of the field energy in **(59)** which neglects per definition the terms linear in $\varphi$ in the mixture rule of the dielectric constant. These terms have to be included in the calculation of the profiles as shown in section 2, eq.**(10)**.

Second, the authors in [11] and [13] include the contribution of vacuum polarization while the calculation here relies on the definitions **(43)**, **(44)** and **(46)**.

Third, equation **(59)** is taken originally from [13]. The dimensions of this equation are corrected with respect to the constant factor $v_m$. The value $v_m$ is taken from from [16] p.797 It should be said that all these changes by the author here do not lead to any changing consequences in the results and conclusions of both references [11] and [13].)

In fig.5 a polymer solution is placed in a cylindrical capacitor as dielectrics. If the capacitance is charged a inhomogeneous electric field arises in the capacitance which entrains a radial profile of the polymer volume fraction $\varphi$ and also a radial profile of the dielectric constant $\varepsilon$.

Because no function is known for the dielectric constant vs. $\varphi$ the most simple non-linear mixture rule is taken to be (k= factor, definition of indices p:=polymer , s:=solvent)

$$\varepsilon = \varphi \varepsilon_p + (1-\varphi) \varepsilon_s + k\varphi(1-\varphi) \tag{63}$$

The density profile can be calculated by application of the equilibrium condition **(6)** $\partial \mu_P^* / \partial r = 0$ to the chemical potential **(58)**. This yields the differential equation of the profile

$$\frac{\partial \varphi}{\partial r} = - \frac{C^2 \varepsilon_0 \beta (\varepsilon_p - \varepsilon_s + k(1-2\varphi(r)))}{\left(\frac{1-2\zeta+\varphi(r)/N+\varphi(r)}{v_m} + \frac{C^2 \beta k \varepsilon_0}{r^2}\right) r^3} \quad (64)$$

with $C$ defined as $C := Q/(2\pi\varepsilon_0 h)$.

Equation **(64)** is solved by the program in appendix 2 for a state above the critical point. The calculated profile is shown in fig.5 . Applying the formula of the cylindric capacitance $C$

$$\frac{1}{C} = \int_{r_1}^{r_2} \frac{dr}{2\pi\varepsilon_0 \varepsilon(r) r h} \quad (65)$$

the capacitance can be evaluated with the data in fig.5 using a tabel calculation program. It is a interesting question whether the tap in the fig.4a can be avoided if the following cycle is proceeded without tap using rectangular electric voltage pulses,comp.fig.6: The cycle of the system starts if the field is switched off. The solution may be equally distributed for the 1-phase system, or may be in the demixed state for the 2-phase system.

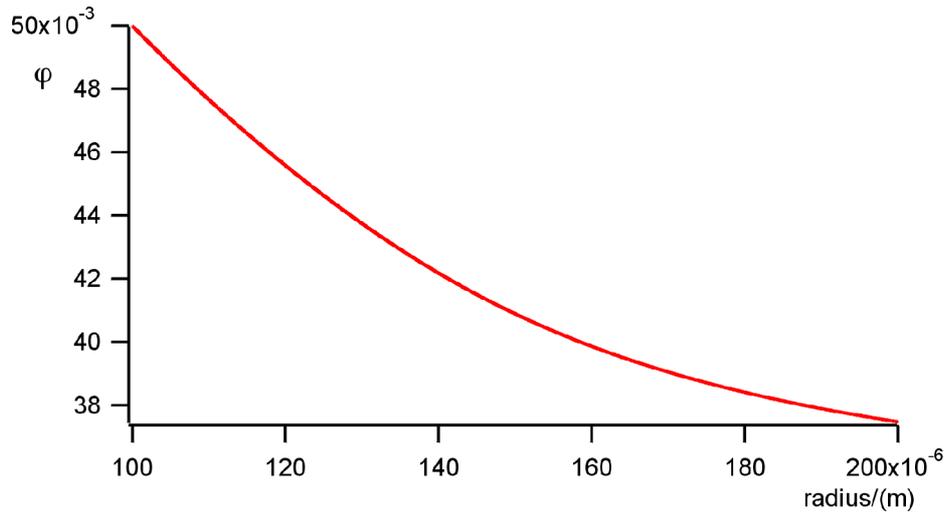

Fig.5: Volume fraction $\varphi$ vs. radius r in a capacitance with cylindric geometry using as dielectrics a overcritical polymer solution polystyrene in cyclohexane . data of calculation: inner diameter 0 .1mm, outer diameter 0.2 mm, height 50 cm, $v_m = 1.53*10^{-28}$, $N = 1/(0.04)^2$, $\zeta = 0.539$, $\epsilon_{solvent} = 5$, $\epsilon_{polymer} = 35$, k =-30, $\beta = 1/(1.38*10^{-23}*300)$, charge $Q = 2.5\times 10^{-8}$, initial value $\varphi(r_1 =.1\text{mm}) = 5\%$

Then, the capacitor is switched on so fast that the diffusion in the solution cannot follow. Therefore, the capacitance is charged at the capacitance *C(U=0)*. Then, if the voltage is high and the voltage remains constant, the solution has time to diffuse and builds up the equilibrium profile at high field. So the capacitance decreases to the equilibrium value *C(U)*, comp. Fig 4d. During this phase a current flows from the capacitance at constant high field. Then, the capacitance *C(U)* is discharged again faster than diffusion and the cycle can start again after the relaxation time necessary for the solution to reach the equilibrium again. Therefore, the cycle here could show an electrically induced "gain hysteresis" due to the relaxation of the dielectric material. It is interesting to ask for the quantitative relevance of this cycle:

A loss was found in the numerical tests performed with the cylindric capacitance shown in fig.5 . However, the parallel geometry with homogeneous field + gravitation, see fig.4a (but without tap) shows the effect principally, even if the effect is negligible quantitatively, as shown in fig.7.

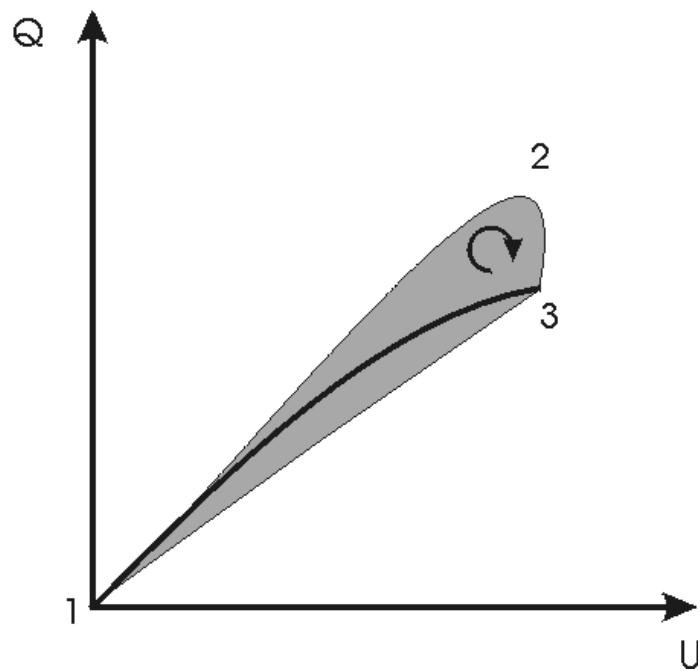

Fig.6: Charge Q vs. voltage U of a gain cycle of a capacitance using a 2-phase polymer solution as dielectrics in a homogeneous electric field driven by rectangular voltage pulses, comp. text
Bold line: equilibrium capacity; weak lines: dynamic non-equilibrium capacities due to retardation by diffusion. The work area is overdrawn, because the predicted gain effects are less than 1 ‰.
1->2 fast charging, 2-3 relaxation of solution with decrease of ε at constant voltage
3->1 fast discharge of capacitance, the after a relaxation time the cycle can start again.

The result is obtained by the following consideration:

Because the gravitational field separates the different phases gravity is included in the ansatz.

Then, the free energy is ($M_{p/s}$:= molar mass of polymer or solvent, $L$:= Avogadro number)

$$f'' = \frac{1}{v_m}\left((\varphi/N)\ln\varphi + \frac{1}{2}(1-2\zeta(T))\varphi^2 + \frac{1}{6}\varphi^3\right) - \frac{\beta}{2}\mathbf{E}^2\varepsilon_0(\varepsilon(\varphi)-1) + \frac{gx\beta}{v_m L}(M_p\varphi + M_s(1-\varphi)) \quad (66)$$

The application of the equilibrium conditions eq. **(6)** ($\mu = \partial f''/\partial\varphi = const.$) leads to the differential equation of the concentration profile of $\varphi(h)$ vs. height h in the gravitational field

$$\frac{\partial\varphi}{\partial h} = -\frac{gN(M_p-M_s)\beta\varphi(h)}{L(1+N\varphi(h)-\varepsilon_0\mathbf{E}^2\beta Nkv_m\varphi(h)-2N\zeta\varphi(h)+N\varphi(h)^2)} \quad (67)$$

This problem is solved for $\mathbf{E}=5*10^7$ V.cm$^{-1}$ and for a volume of 10 cm height. The profile of the volume fraction $\varphi$ of the polymers between top and down is linear with height and shows ~ 2‰ deviation per 10 cm at the chosen total concentration of $\varphi=\varphi_c=4\%$ above the critical point($\zeta=.51<\zeta_c$). This means that the influence of gravitation can be neglected against the influence of strong electrical fields for a 1- phase system.

In a 2-phase solution the higher concentrated gel phase is the sediment, because the monomer ($M_{styrene}$ = 104.144 g.mol$^{-1}$) weights more than the solute ($M_s = M_{cyclohexane}$=84.162 g.mol$^{-1}$) while the specific volume of both components are nearly the same($v_{cyclohexane}$= 9.256x10$^{-5}$ m$^3$.kmol$^{-1}$ [17], $v_{styrene} = v_m L$= 9.2139x10$^{-5}$ m$^3$.kmol$^{-1}$ and $v_m$=1.53x10$^{-28}$ m$^3$ ).

The estimation of the gain of a complete electrical cycle is done under the following conditions:

If the solution is under the field **E** it is set always exactly at the critical point and mixes. If then the field is switched off the same solution changes to the demixed state in the two-phase area.

Applying the conditions of a critical point on **(61)** it follows with **(59)** and **(63)**

$$\text{from } \frac{d^2\mu^+}{d\varphi^2}=0: \qquad \varphi_C = N^{-1/2}$$

$$\text{from } \frac{d\mu^+}{d\varphi}=0: \qquad \zeta_C(\mathbf{E}) = \zeta_C(\mathbf{E}=0) - \frac{1}{2}\varepsilon_0 \beta \mathbf{E}^2 v_m k \qquad (68)$$

using the abbreviation $\zeta_C(\mathbf{E}=0) = 0.5 + N^{-1/2}$.

The second equation allows to calculate a corresponding $\zeta(T) = \zeta_C(\mathbf{E})$ for the demixed and field free state of every corresponding field **E** applied at the critical point.

Now, by applying a known parameter representation of the phase separation line the volume fractions of the sol $\varphi'$ and the gel phase $\varphi''$ are calculated by a graphical method using the source code in appendix 3. The equations of the parameter representation without field are [16]

$$\varphi'' = t\varphi'$$

$$\sqrt{N}\varphi' = \sqrt{\frac{6(t+1)\ln(t) - 12(t-1)}{(t-1)^3}} \qquad (69)$$

$$(1/2 - \zeta)\sqrt{N} = \frac{2\ln(t)(t^2+t+1) - 3(t^2-1)}{\sqrt{[6(t+1)\ln(t) - 12(t-1)](t-1)^3}}$$

The solutions $\varphi'$ and $\varphi''$ from **(69)** determine the dielectric constants of the 2-phase system.

$$\varepsilon' = \varepsilon_p \varphi' + (1-\varphi')\varepsilon_s + k\varphi'(1-\varphi')$$
$$\varepsilon'' = \varepsilon_p \varphi'' + (1-\varphi'')\varepsilon_s + k\varphi''(1-\varphi'') \qquad (70)$$

The splitting factor $S^*$ can be calculated by solving the relation

$$\varphi_C = \varphi' S^* + (1 - S^*)\varphi'' \qquad (71)$$

for $S^*$. This allows to calculate the combined effective dielectric constant of the 2-phase system.

$$\varepsilon = \varepsilon' S^* + (1-S^*) \varepsilon'' \tag{72}$$

This is compared against the mixed dielectric constant $\varepsilon_c$ at the critical point which is

$$\varepsilon_c = \varepsilon_p \varphi_c + (1-\varphi_c) \varepsilon_s + k \varphi_c (1-\varphi_c) \tag{73}$$

Therefrom, the gain efficiency $\eta$ of the electric cycle can be estimated acc. to

$$\eta := \frac{\Delta E}{E} \approx \frac{\varepsilon_c - \varepsilon}{\varepsilon_c - 1} \tag{74}$$

where $\Delta E$ is the working area in fig.6 approximated as a triangle and E is the energy supplied. Fig. 7 shows the estimated gain efficiency vs. the electric field strength. It is clear that any measurable gain effects are beyond the breakdown voltage of the material under discussion. Therefore, in effect, no energy gain is measurable. However, from the general theory it can be questioned whether this holds generally if other materials, suited chemical equilibria or electric double layers in plasmas, liquids and solids are used as dielectrics in optimized geometries. Analogous considerations with ferrofluids in magnetic fields could also be possible.

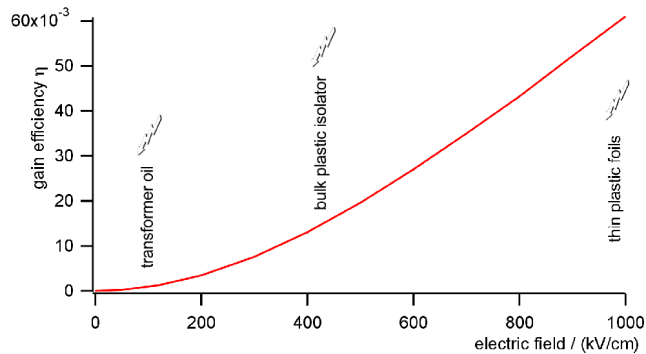

Fig.7: calculated gain efficiency per cycle vs. maximum voltage of the polystyrene/cyclohexane model
inserts show breakdown voltages of different isolation materials for comparison, comp.text.
data of calculation: $v_m = 1.53*10^{-28}$, $N = 1/(0.04)^2$, $\epsilon_{solvent} = 5$, $\epsilon_{polymer} = 35$, k =-30,
$\beta = 1/(1.38*10^{-23} *300)$, $\varphi=\varphi_c =4\%$

**Tab.1:** analogous features between thermodynamics and mechanics

| mechanics | thermodynamics |
|---|---|
| time mean or least action functional | ensemble average |
| Hamilton energy | Hamilton energy |
| non-extremal state of functional | non-equilibrium state |
| Legendre transformations, i.e. L, H | Legendre transformations, i.e. U, H, F, G |
| Pontrjagin's extremum principle | extremum principle of potentials |
| second variation of the Hamiltonian | "second law" |

*5.Conclusion*

The analogies between mechanics and thermodynamics- shown in tab.1- suggest that the direction of irreversibilities can be understood from a variation principle applied to the Hamilton energy analogously to the extremum principle of Pontrjagin applied to mechanics. Therefrom, it can be derived the grand partition distribution, the maximum entropy principle and the second law in a modified form which can violate Clausius's integral version if cycles with changing potential fields are included into consideration. It was shown that this is not in contradiction to the maximum entropy principle which also follows from the initial ansatz.

Acc. to a conventional view the described contradictions to second law would be due to a forbidden "strange" material behaviour [18] due to a wrong or insufficient theoretical model. Acc. to the purely mathematical approach of thermodynamics- presented here- strange material behaviour is excluded a priori. All empirical information about a system is in the thermodynamic potential describing the material behaviour. Therefrom, the application of all possible variation principles allows to determine the directions of the irreversible processes. Some experimental data speak in favour for the purely mathematical approach.

Therefore, all systems which show this theoretical contradiction between second law and the second variation of Hamilton energy could be interesting for further research.

**Note added in proof from 28$^{th}$ October 2003:**

In the meantime experimental data of a analogous system with ferrofluids in magnetic fields [18][19] are published which at first sight seem to sight contradict to the results of the calculation in this article, because they show an instability of the solution with a increasing magnetic field. The corresponding calculations of these system [18] may be the current state of the art today and seems to explain the data approximately even quantitatively, however, beside the known problem of polydispersity, some deficiencies in the calculation make it difficult to determine exactly whether there is a loss or a gain in a cycle with ferrofluids. The following critical points have been found:

1) The boundary condition at the phase boundary of a 2-phase ferrofluid system is not accounted for. At the boundary holds $\mathbf{B_1} = \mathbf{B_2}$ for phase 1 and 2 due to $\nabla \cdot \mathbf{B} = 0$. Because $\mu_1 \neq \mu_2$ holds follows $\mathbf{H_1} \neq \mathbf{H_2}$. . The authors in [18] assume $\mathbf{H_1} = \mathbf{H_2}$. This can influence or even reverse the result of the whole calculation.

2) A constant homogeneous field in the solution can be assumed in the solution only for special bowl geometries [20]. They are not realized in the setup of [18]. The author do not discuss either the influence either any space dependence of the field energy in the solution and therefore overlook or neglect any gradients in concentration and pressure in the solution.

3) The authors in [14] calculate the phase equilibrium assuming same osmotic pressure equilibrium in both phases. This is surely qualitatively wrong because their calculation neglects any force due to the field pressure terms which arise at the phase boundary due to the spatial changing magnetic field energy. The use of a Maxwell construction [16] would avoid this mistake.

The differences of the ferrofluid system with the system in this preprint point out to the fact that the cited model of section 4 shown here and as well the models of [18,19] are insufficient (In section 4 they may be thermodynamically inconsistent). We sketch here only how to settle the question but do not solve it here:

For the ferrofluid system one has to solve the partial equation system

$$\Delta \Phi_{\mathbf{B}}(x_i(r), r) = 0$$
$$\frac{\partial \mu_i^*}{\partial r}(x_i(r), \mathbf{B}(r)) = 0 \qquad (75)$$

where $\Phi_{\mathbf{B}}$ is the magnetostatic potential and **B** is the magnetic field.

For the electrostatic system holds

$$\Delta \Phi_{\mathbf{D}}(x_i(r), r) = -4\pi \rho_E$$
$$\frac{\partial \mu_i^*}{\partial r}(x_i(r), \mathbf{E}(r)) = 0 \qquad (76)$$

where $\rho$ represents the charge distribution, which generates the field.

A look on the total change of field energy in the volume will answer the question and will show whether energy flows in or out of a capacitance or coil during a mixing process and will answer the question for the system discussed above.

**Appendix 1:** Algorithm to solve the phase equilibrium of mixture in a field, comp.section 2

Problem:

A volume containing a mixture Argon-Methane is rotated. The inner rim of the volume is at $r_1$, the outer rim at $r_2$, the cross section is constant. Without field applied the volume is filled with a mixture of molar ratio $x_i$, spec. volume $v$ at temperature $T$.

Under the influence of the centrifugal field the mixture distributes inhomogeneously in the volume. The distribution of spec. volume $v$, molar ratio $x_i$ and pressure $P$ has to be calculated.

Solution:

In order to solve the problem the following subroutines are written which are listed here from lowest to highest level. All iterating subroutines use the Newton-Raphson-technique.

**PMu_VX** :  calculates the equations of state, $P=P(v,x_i; T)$, $\mu_i=\mu_i(v,x_i; T)$
We use a Bender equation of state (EOS) [19] with the material data:

|  | Argon | Methane |
|---|---|---|
| mol. weight | 39.948 | 16.043 |
| crit. pressure/(Pa) | 4865300 | 4598800 |
| crit. volume/(m³/mol) | 7.452985075e-5 | 9.9030865D-05 |
| crit. temperature/(K) | 150.69 | 190.56 |
| Omega | -.00234 | .0086 |
| Stiel factor | .004493 | .00539 |

fit constants of mixture: $k_{ij}$ = .9977865068023702   $Chi_{ij}$ =1.033181352446963
$Eta_M$ =2.546215505163194

**V_PX**:  inverts the EOS **PMu_VX** by solving $P_0 - P(v; x_i, T) = 0$ for $v$ numerically

**VX_Mu**:  inverts the EOS **PMu_VX** by solving $\mu_i^0 - \mu_i(v, x_i; T) = 0$ for $(v,x_i)$ numerically

**VOLUME**: solves the complete thermodynamic state in a volume acc. to the method described in the example of section 2. The algorithm proceeds as follows:

- Define the number m of volume array cell, i.e. the partition of the volume.
- Give pressure $P_0$, temperature $T_0$ and concentrations $x_i$ at one point $r_{ref}$ called the reference point
- Using the subroutine **V_PX** invert numerically the equation of state at $r_{ref}$ and calculate v
- Initialize the numbers of all particles in the volume, i.e. $N_i = 0$
- FROM volume array cell j = 1 TO m
  - Calculate all interesting thermodynamic data at $r^j$ using the EOS **PMu_VX**
  - Calculate $dn_i(r^j)$, i.e. the number of particles of each sort i in this subsection $dV(r_j)$
  - $N_i = N_i + dN_i(r_j)$
  
  Add up the particle number in the compartment to the total number of the array
  Calculate $P^{j+1}$ and $\mu_i^{j+1}$ in the adjacent volume section $dV_{j+1}$ acc. to **(15,16)**.
  - Invert numerically the equation of state at $r_{j+1}$ and calculate v,,$x_i$ using the subroutine **VX_Mu**
- NEXT J
- Give out all calculated values $P^j, v^j, x_i^j, \mu_i^j$.

**JACOBIMATRIX**: calculates numerically all derivatives and vectors of the Jacobimatrix necessary to solve the posed problem by the Newton-Raphson procedure

**GAUSSALGORITHM**: solves a linear system of equations

**N_PX**: calculates the reference values $P^{ref}$ and $x_i^{ref}$ for the volume array under the constraint of mass conservation, meaning $N_i^0 = constant$.

The basic idea of the main program routine **N_PX** is the following:

The total numbers $N_i$ of particles calculated by the subroutine **VOLUME** are regarded as numerical functions of the starting values $P_{ref}$ and $x_i^{ref}$

$$N_i(P_{ref}, x_i^{ref}) = \sum_{j=1}^{m-1} \frac{x_i(r_j) A(r_j)}{v(r_j)} \Delta r \qquad (77)$$

In order to find the correct starting values $P_{ref}$ and $x_i^{ref}$ for the given particle numbers $N_i^0$ the following equation has to be solved numerically for $\vec{X} = (P_{ref}, x_i^{ref})$

$$\Delta\vec{N}(\vec{X}) = N_i(P_{ref}, x_1^{ref}...x_{n-1}^{ref}..) - N_i^0 = 0 \qquad (i = 1, 2, ...n) \quad (78)$$

This is done in two steps: First, the (subroutine) **JACOBIMATRIX** $\bar{\bar{J}}$ has to be calculated which calls the subrotine **VOLUME** repetively in order to calculate the $N_i$ necessary for the numerical derivatives and the difference vector $\Delta\vec{N}(\vec{X})$ of **JACOBIMATRIX**.. Then, (the) **GAUSSALGORITHM** solves

$$\bar{\bar{J}} \cdot \Delta\vec{X} = \Delta\vec{N}(\vec{X}) \qquad (79)$$

for the step width $\Delta\vec{X}$ to the next point of iteration using the data calculated by (the) **JACOBIMATRIX**. If the accuracy of the calculation is too low the iteration restarts at the next point of iteration. If the accuracy is high enough the iteration is stopped and the calculated reference values are used to determine the full thermodynamic state calling the subroutine **VOLUME**.

The principal structure of the whole program above is shown in fig.8 .

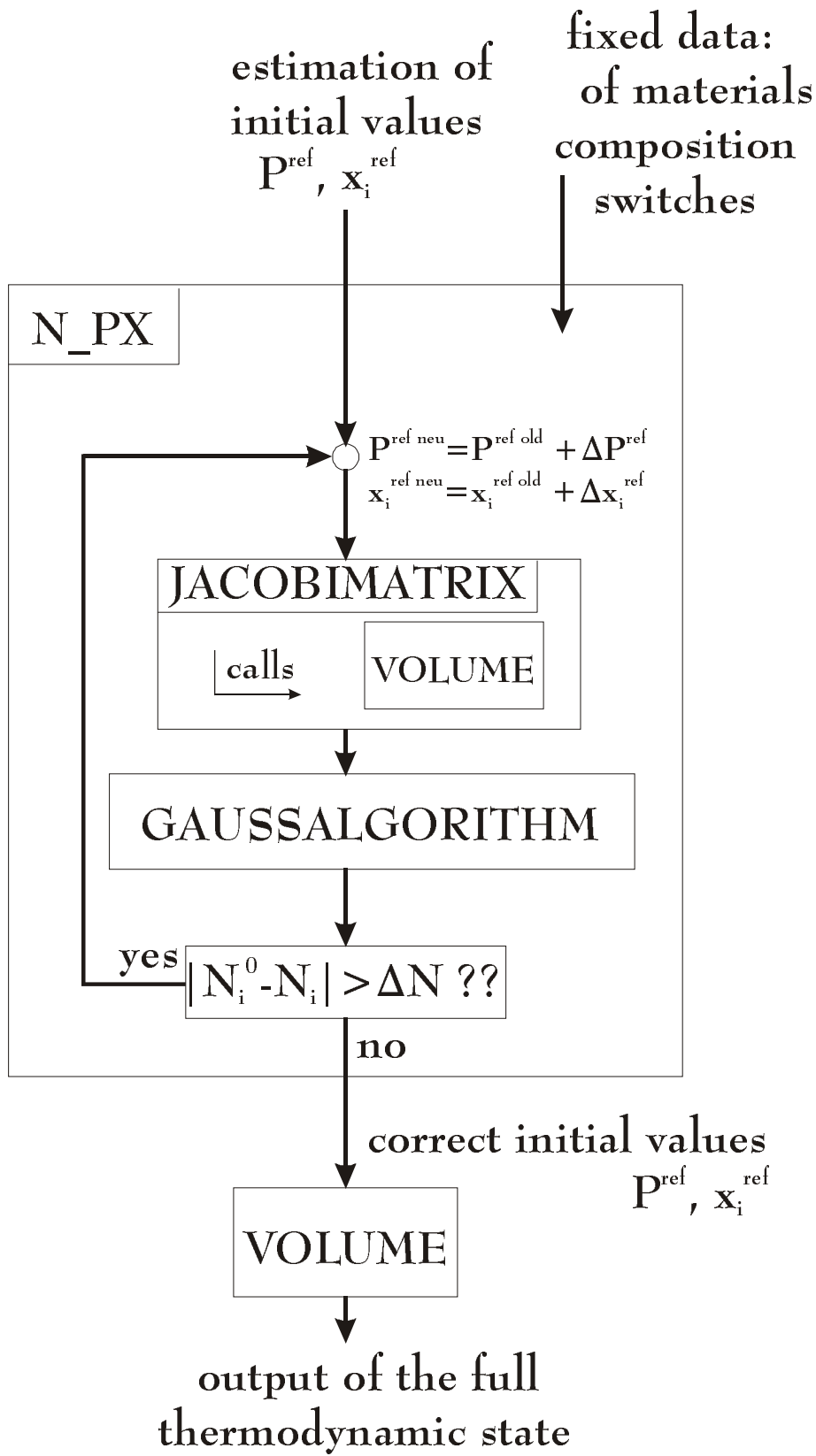

Fig.8: The principal program structure for solving the problem from appendix 1, comp.text.

**Appendix 2:** The MATHEMATICA source code generating the profile in fig.5

```
eps[φ[r]] =   epspolminusone*φ[r] + (1 - φ[r])*epssolvminusone + k*φ[r]*(1 - φ[r])
ELF[r] = Const/ r
μ[r] = (1/N +  Log[φ[r]]/ N + (1 - 2*ζ)*φ[r] + φ[r]² /2)/vm
       + β/2*diel*(ELF[r])²  * D[eps[φ[r]],φ[r]]
Equation = D[μ[r], r]
Solve[Equation == 0, φ'[r]]

 diel = 8.854*10⁻¹²
 vm = 1.53*10⁻²⁸
 N = 1/(0.04)²
 ζ = 0.539
 epssolvminusone = 4
 epspolminusone = 34
 k =-30
 β = 1/(1.38*10⁻²³ *300)
 Q = 2.5 *10⁻⁸
 start = 0.0001
 end = 0.0002
 h = 0.5
 Const =Q/(2*Pi*diel*h)
 Eschaetz = 2 * Const/(start + end)
 NDSolve[{φ'[r] ==-(Const² *diel*β*(-epssolvminusone + epspolminusone +
           k*(1 - 2*φ[r])))/(r³ *(Const² *diel*β/r² +(1-2*ζ+1/N+φ[r])/vm), φ[start] == 0.05},
{φ},{r,start,end}]
 Plot[Evaluate[φ[r] /. %], {r, start, end}]
 phi[r_] := Evaluate[φ[r] /. %%]

n = 10000
phi[start]
Arr = Table[phi[start + i * (end - start)/n], {i, 0, n, 1}];
Arr >> Phi.txt
```

**Appendix 3:** The MATHEMATICA-code generating graphical solutions for calculating fig.7

```
diel = 8.854*10^-12
β = 1/(1.38*10^-23 *300)
φc = .04
vm = 1.53 10^-28
epspolminusone = 34
epssolvminusone = 4
k = -30
epsnull = φc*epspolminusone + (1 - φc)*epssolvminusone + k*(1 - φc)*φc
El = 1.2*10^7
Δζ = -diel*β*El^2 *k*vm/2
ζ = .54 + Δζ
N = 625
Y0[t_] := (.5 - ζ)N^.5 + ((2*(t*t + t + 1)*Log[t] -
     3*(t*t - 1))/((6*(t + 1)*Log[t] - 12*(t - 1))^.5*(t - 1)^1.5))
ParametricPlot[{X[t], Y0[t]}, {t, 1.5, 2}]
ParametricPlot[{t*X[t], Y0[t]}, {t, 1.5, 2}]

X1 = .7045
X2 = 1.359
T = X2/X1
φ1 = X1*φc
φ2 = X2*φc
S = (φc - φ2)/(φ1 - φ2)
eps1 = φ1*epspolminusone + (1 - φ1)*epssolvminusone + k*φ1*(1 - φ1)
eps2 = φ2*epspolminusone + (1 - φ2)*epssolvminusone + k*φ2*(1 - φ2)
eps = S*eps1 + (1 - S)*eps2
gain = (eps - epsnull)/epsnull
```